\documentclass[balancelastpage,aps,pra,twocolumn,eqsecnum,superscriptaddress,showpacs,floatfix,nofootinbib]{revtex4}
\usepackage{graphicx}
\usepackage{amsmath,amssymb}
\usepackage{bm}

\def\lan{\langle}
\def\ran{\rangle}

\newcommand{\bra}[1]{\mbox{$\left\langle #1 \right|$}}
\newcommand{\ket}[1]{\mbox{$\left|#1\right\rangle$}}
\newcommand{\braket}[2]{\mbox{$\langle #1  | #2 \rangle$}}
\newcommand{\Mod}[1]{\mbox{$\left|#1\right|$}}

\def\d{\displaystyle}

\newlength{\cdgap}
\def\el{\\[\cdgap]}

\def\p{\partial}
\def\tr{\mbox{tr}}

  \def\O{{\cal O}}
\def\S{\mbox{$\cal S$}}

\def\al{\alpha}  
  
 \def\la{\mbox{$\lambda$}} \def\si{\sigma}

\def\ua{\uparrow} \def\da{\downarrow}

\def\1/2{\frac{1}{2}} \def\1/3{\frac{1}{3}} \def\3/2{\frac{3}{2}}

\begin{document}
\rule{1cm}{0mm}\title{Chiral phase from three spin
interactions in an optical lattice}
\author{Christian D'Cruz}
\affiliation{Department of Mathematics, Royal Holloway, University
of London, Egham, Surrey TW20 0EX, UK}
\email{C.H.D-Cruz@rhul.ac.uk}
\author{Jiannis K. Pachos}
\affiliation{Department of Applied Mathematics and Theoretical
Physics, University of Cambridge, Cambridge CB3 0WA, UK}
\email{j.pachos@damtp.cam.ac.uk}

\date{\today}

\begin{abstract}

A spin-1/2 chain model that includes three spin interactions can effectively
describe the dynamics of two species of bosons trapped in an optical lattice
with a triangular-ladder configuration. A perturbative theoretical approach
and numerical study of its ground state is performed that reveals a rich
variety of phases and criticalities. We identify phases with periodicity
one, two or three, as well as critical points that belong in the same
universality class as the Ising or the three-state Potts model. We establish
a range of parameters, corresponding to a large degeneracy present between
phases with period 2 and 3, that nests a gapless incommensurate chiral
phase.

\end{abstract}


\maketitle

\section{Introduction}

The proposal \cite{Jaks98} and subsequent realization
\cite{Raithel,Mandel1,Mandel3} of optical lattices for the
manipulation of ultra-cold atoms has attracted a great deal of
research towards the implementation of quantum computation
\cite{Deutsch,Jaksch,Ripoll} and the simulation of condensed matter
systems \cite{Kuklov,Belen,Kukl,Jask03,Duan}. The main advantages of optical
lattices are their long decoherence times and high degrees of
controllability. Hence, one is able to probe phenomena that manifest
themselves at higher orders in perturbation theory such as many body
interactions \cite{Pachos}. This gives the possibility to engineer and
control exotic interactions in many body systems and subsequently realize
novel phases of matter. Examples include cases where frustration effects are
present or competing phenomena coexist giving rise to large degeneracy
structures \cite{Sachdev,Nielsen}. Towards this direction we consider a
semi-one dimensional lattice comprising a triangular ladder. The dynamics of
two boson species mounting the optical lattice in the limit of strong
collisional interactions can be effectively described by a chain of
spin-$1/2$ interacting particles.  Here we consider systems with two and
three spin interactions given by $ZZ=\sum_i \sigma_i^z
\sigma_{i+1}^z$ and $ZZZ=\sum_i\sigma_i^z \sigma_{i+1}^z \sigma_{i+2}^z$
respectively.  We shall see explicitly that it is possible to make $ZZZ$
dominant by appropriate tuning of the collisional and tunnelling couplings.
As simple as these interactions may seem, when combined, they give rise to a
rich variety of phase transitions, the study of which is the subject of this
article.

\begin{center}
\begin{figure}[!h]
\includegraphics[width=0.45\textwidth]{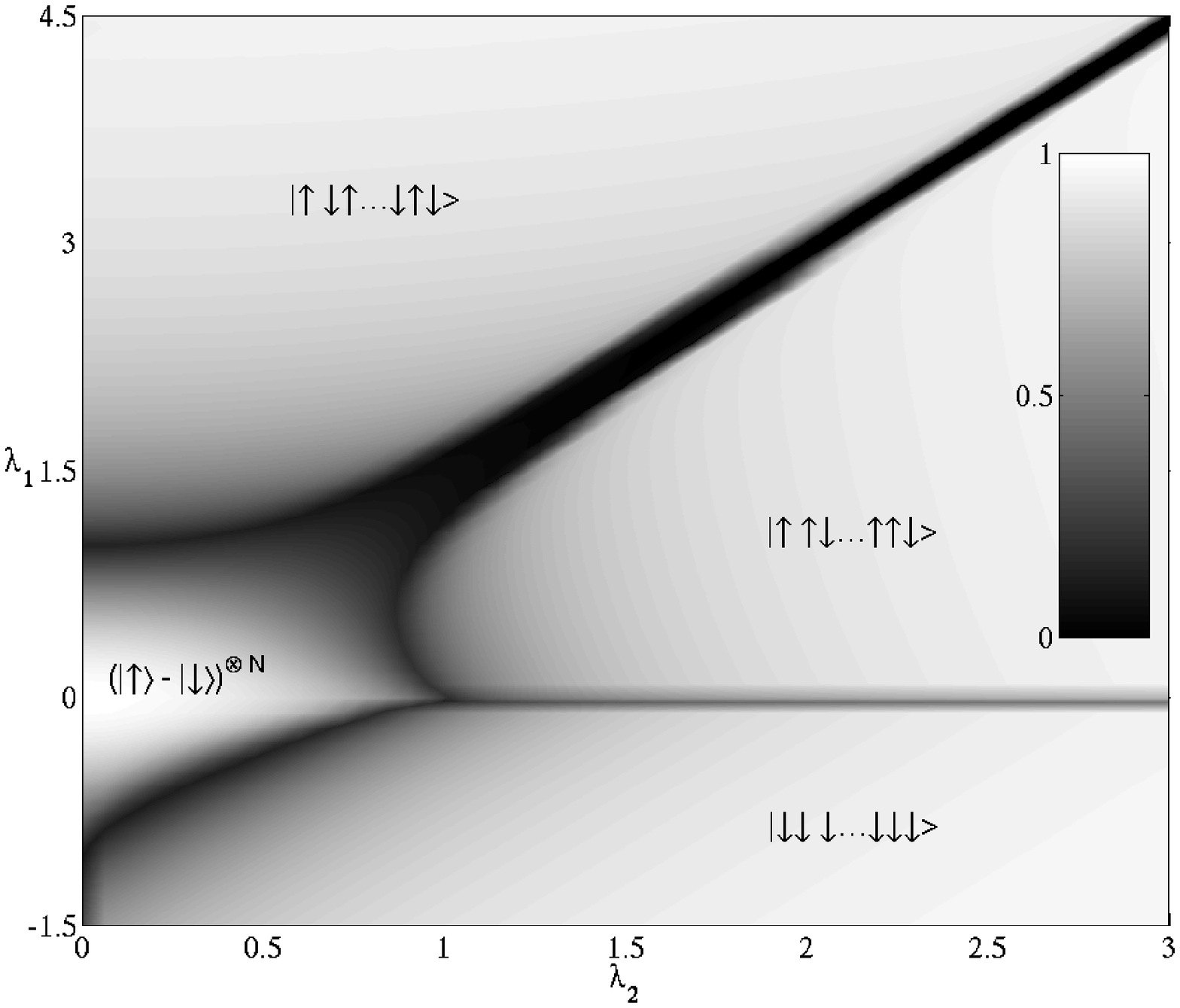}
\caption{\addtolength\baselineskip{0.5mm} A plot of the fidelity,
$\Mod{\braket{\psi}{\phi}}^2$, of the ground state \ket{\psi} with
the predictions \ket{\phi} obtained from each individual term of the
Hamiltonian. The state $\ket{\phi}=\ket{\ua\da\ua\ldots\da\ua\da}$
corresponds to $ZZ$, $\ket{\phi}=\ket{\ua\ua\da\ldots\ua\ua\da}$
corresponds to $ZZZ$ and $\ket{\phi} = (\ket{\ua} -
\ket{\da})^{\otimes N}$ corresponds to the transverse field.  The
state $\ket{\phi}=\ket{\da\da\da\ldots\da\da\da}$ is the common
ground state of $ZZ$ and $ZZZ$ for this range of couplings.  The
spin chain is of length $N=18$.} \label{fig:fidel}
\end{figure}
\end{center}

In particular, we consider a Hamiltonian that includes the two and three
spin interactions with couplings $\la_1$ and $\la_2$ respectively, in the
presence of a transverse field with unit amplitude. We observe that for
$\la_2=0$ the model is exactly solvable, as it reduces to the Ising
interaction with critical points at $\lambda_1=\pm 1$. For $\la_1=0$ we
obtain a Hamiltonian that is not analytically diagonalizable. Nevertheless,
a numerical study reveals a tricritical point at $\lambda_2=1$ that is in
the same universality class as the three-state Potts model
\cite{Wu}. When the two spin interaction is dominant the ground
state has a spin-order of period 2 while the dominance of the higher order
interaction brings about a spin-order of period 3. Though these phases are
gapped, there are values of relative couplings where these orders compete,
giving rise to a high degree of degeneracy. This characteristic allows for
the presence of a gapless incommensurate chiral phase that extends to a wide
range of parameters. In Figure \ref{fig:fidel} the fidelity of the actual
ground state of the Hamiltonian is plotted against the ground state of each
individual term comprising it.  While we see that these agree for a large
range of couplings there is an area between spin-order 2 and 3 where these
states fail to accurately describe the nature of the system. We shall
identify this as the gapless incommensurate chiral phase.  A similar region
has been presented by Fendley, Sengupta and Sachdev \cite{Sachdev} in a
1-dimensional hard boson system.

The article is organized as follows. In Section \ref{optical} we
present the realization of the two and three spin interactions by
ultra-cold atoms superposed by optical lattices. Section
\ref{sec:tricritical} is a study of the tricritical point present at
$\la_1 = 0, \, \la_2 = 1$.  In Section \ref{sec:general}, we present
the properties of the full Hamiltonian. For that we employ
perturbation theory, a bosonization of the model and a study of an
effective field theory that eventually reveal the incommensurate
chiral phase. Finally, in Section \ref{Conclusion}, we conclude and
present future directions.

\begin{center}
\begin{figure}[!b]
\resizebox{!}{2cm} {\includegraphics{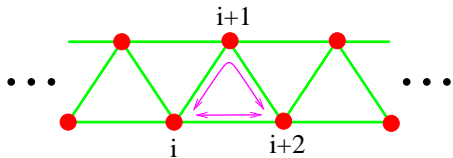} } \caption{\label{chain} The one dimensional chain constructed
    out of equilateral triangles. Three-spin interaction terms appear, e.g. between
    sites $i$, $i+1$ and $i+2$ as for example tunnelling between $i$ and $i+2$
    can happen through two different paths, directly and through site
    $i+1$, the latter resulting into an exchange interaction between
    $i$ and $i+2$ that is influenced by the state of the site $i+1$.}
\end{figure}
\end{center}
\begin{center}
\begin{figure}[!t]
\resizebox{!}{6.0cm}{\includegraphics{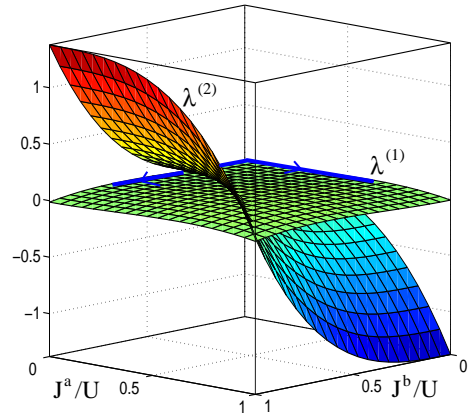}} \caption{\label{comb2} The effective couplings $\lambda_{1}$ and
  $\lambda_{2}$ are plotted against $J^a/U$ and $J^b/U$ for
  $U_{aa}=U_{bb}=2.12U$ and $U_{ab}=U$. The coupling $\lambda_{1}$
  appears almost constant and zero as the unequal collisional
  couplings can create a plateau area for a wide range of the
  tunnelling couplings, while  $\lambda_{2}$ can be varied freely to
  positive or negative values.}
\end{figure}
\end{center}

\vspace{-2.7cm}
\section{Optical Lattices And Three Spin Interactions}\label{optical}

Consider the physical setup where an ultra-cold atomic cloud of two
different species, $a$ and $b$, is superposed with a three dimensional
optical lattice. The atoms can tunnel through the potential barriers of the
lattice from one site to the next with a coupling $J$. When two or more
atoms are present in the same site, they collide with coupling $U$. For
sufficiently large intensities of the laser radiation, where $J\ll U$, the
system is in the Mott insulator regime with a regular number of atoms per
site. In particular, we can arrange the density of the atomic cloud to be
low enough so that only one atom can exist at each site of the lattice, i.e.
$\langle n_a + n_b\rangle \approx 1$. In this way each lattice site is a
simple two state system, that can be viewed as a spin-1/2 particle. Within
this representation, the Hamiltonian can be written in terms of Pauli spin
operators.

Indeed, for the triangular ladder seen in Figure \ref{chain} we obtain the
Hamiltonian of the form \cite{Pachos,Pachos1}
\begin{equation}
H = B \sum \sigma_i^x  + \lambda_1 \sum \sigma_i^z \sigma_{i+1}^z +
\lambda_2 \sum \sigma_i^z \sigma_{i+1}^z \sigma_{i+2}^z,
\label{Ham}
\end{equation}
where $B$, $\lambda_1$ and $\lambda_2$ are all functions of the initial
tunnelling and collisional couplings, $J$ and $U$.  The values of
$\lambda_1$ and $\lambda_2$ can be controlled independently by varying $J$
and $U$.  In particular, it is possible to make $\la_2$ large compared to
$\la_1$ as can be seen in Figure
\ref{comb2}. A careful consideration of the system would reveal that
terms of the form $\sigma^z_i \sigma^z_{i+2}$ also appear, due to
the triangular configuration.  This can be remedied by employing
superlattices that do not alter the other terms.

\section{The $ZZZ$ Interaction}\label{sec:tricritical}

Ultimately, we would like to study the entire $(\lambda_1,
\lambda_2)$ plane, identify the regions of criticality behavior and
understand the global properties of the ground state. Before turning to the
general problem let us consider the two special limiting cases. Initially,
when $\la_2=0$, the Hamiltonian reduces to the Ising interaction between
neighboring spins in the presence of a transverse magnetic field. This
well-studied model exhibits criticality behavior for $\lambda_1=\pm 1$.

Another interesting model \cite{Alcaraz} can be obtained for $\la_1=0$. For
simplicity we can take $\la_2 = 1$ and consider the Hamiltonian to
be a function of the transverse field with amplitude $B$
\cite{Penson,Igloi}. As we vary $B$ we observe that there are two
distinctive regions. For $B \gg 1$, the ground state has all the spins
oriented towards the $x$ direction, and for $B\ll 1$ there is a degeneracy
between the states $\ket{\ua\ua \da \ua \ua \da\ua
\ua\da}$, its translations (the $Z_3$ symmetry) and
$\ket{\da\da\da\da\da\da\da\da\da}$. It is of interest to study the
behavior of the system between these two limiting cases.

Due to the symmetry of the Hamiltonian we can pinpoint where a
possible critical point can lie. For that, one can define, $ \mu_i^x
\equiv \sigma^z_i\sigma^z_{i+1}\sigma^z_{i+2}$ and $ \mu_i^z \equiv
\prod_{n=0}^{\infty} \si_{i-3n}^x \si_{i-3n-1}^x$. It is easily
verified \cite{Turban} that these operators obey the Pauli algebra
commutation relations. Moreover, under this transformation the
Hamiltonian becomes $B H(B^{-1})$ which has exactly the same
spectrum as $H(B)$. Hence, if there exists a single critical point
it has to be at $B=1$.

\begin{center}
\begin{figure}[ht]
\resizebox{0.48\textwidth}{!}
{\includegraphics[height=15cm]{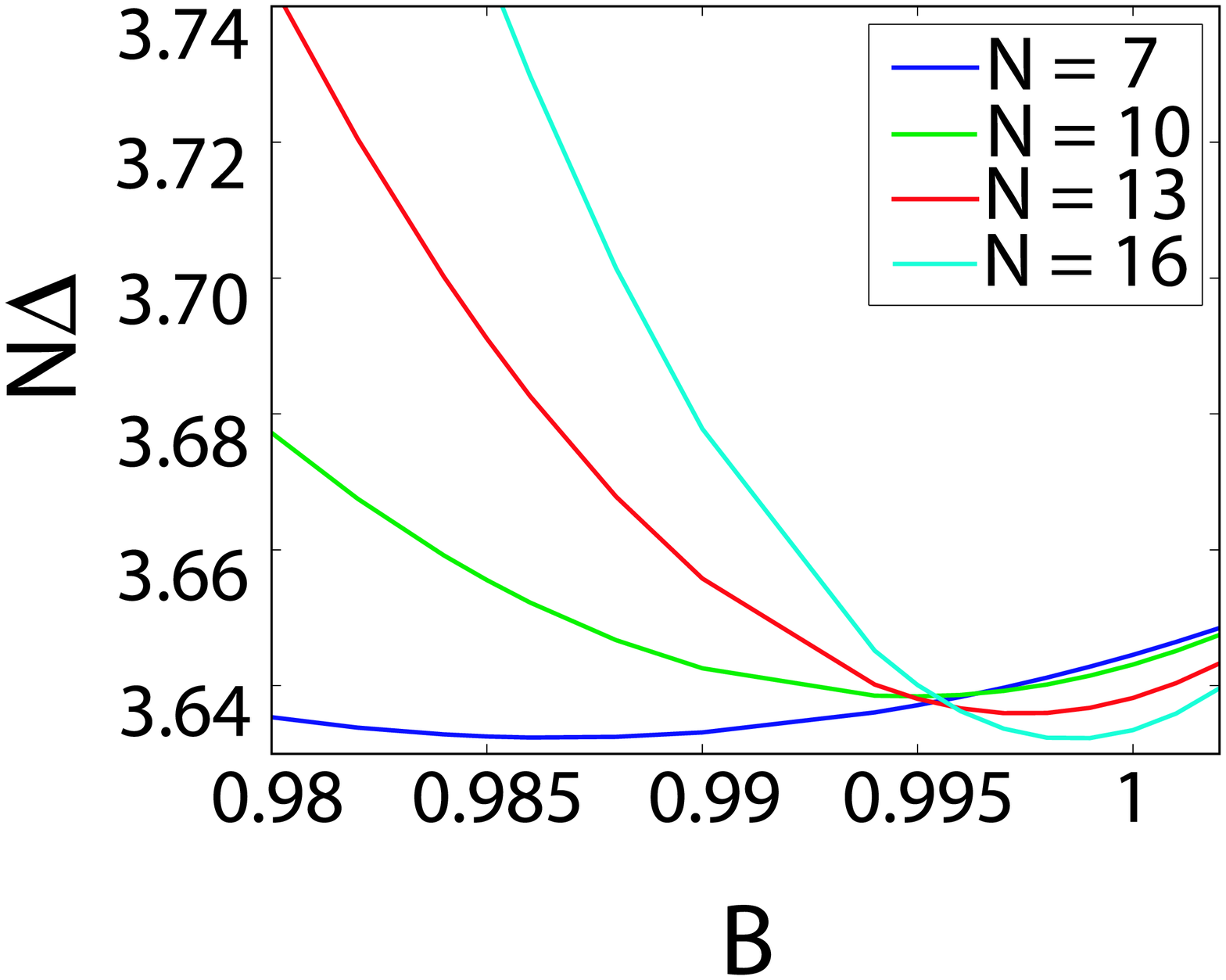} \hspace{1cm}
\includegraphics[height=15cm]{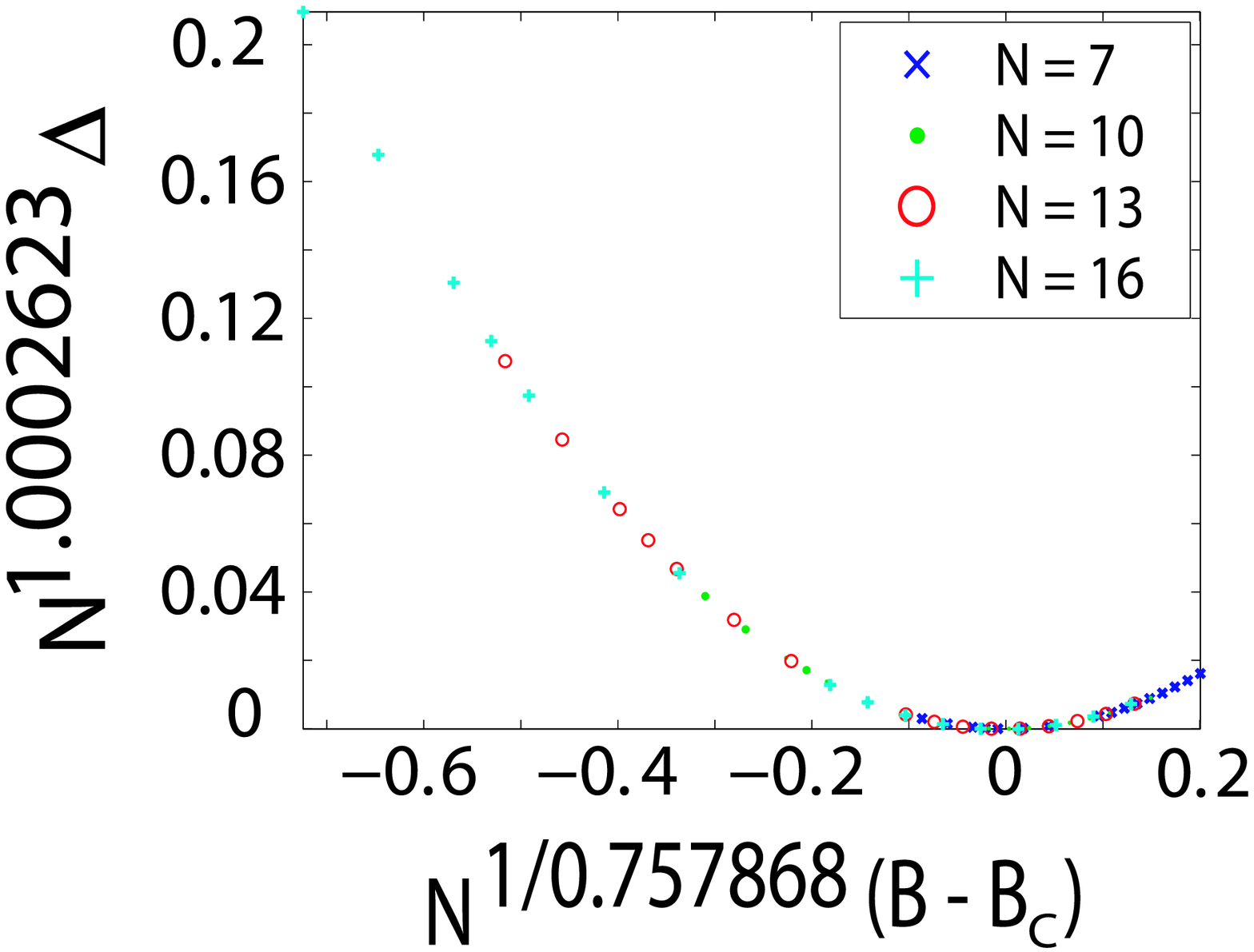} }
\caption{\label{ZZZ1mod3_1} (a) Quantum critical point
signified by the minima of the energy gap $\Delta$. Due to low $N$,
these provide only an estimate for the position of the critical
point. The progression towards $B_x = 1$ as $N$ increases supports
the theoretical predictions. (b) To determine the critical exponents
we have chosen the values of $z$ and $\nu$ that best superpose the
curves of Figure \ref{ZZZ1mod3_1}(a) into one uniform function
$\Phi$.}
\end{figure}
\end{center}

We can verify this numerically and identify the critical exponents, by
observing the minimum in the energy gap between the ground and first excited
state on a finite chain of spins with periodic boundary conditions. In the
thermodynamical limit this minimum, if it corresponds to a critical point,
will become zero. Near this region the energy gap, $\Delta$, is expected to
scale as follows
\cite{Goldenfield},
\begin{equation}
\Delta = N^{-z} \Phi \left(N^{1/\nu}(B - B_c)\right). \label{eqn:Delta}
\end{equation}
Here $\Phi$ is a universal scaling function, $z$ is the dynamic critical
exponent, and $\nu$ is the correlation length exponent. Figure
\ref{ZZZ1mod3_1}(a) gives a plot of $N^{1.00263}
\Delta$ versus $B$, which shows that the energy minimum progresses
towards $B = 1$ for increasing $N$. We may employ Eqn. (\ref{eqn:Delta}) to
estimate $z$ and $\nu$. Figure
\ref{ZZZ1mod3_1}(b) shows $N^{1.0002623} \Delta$ against
$N^{1/0.757868} (B_x - B_x^c)$, where the data for systems of
different sizes come together into a single curve. From Eqn.
(\ref{eqn:Delta}) we deduce the value of the critical exponents to
be $z\approx 1$ and $\nu \approx 0.76$.

\vspace{0.2cm}
\vspace{-0.8cm}

\begin{center}
\begin{figure}[!ht]
\includegraphics[width=0.35\textwidth]{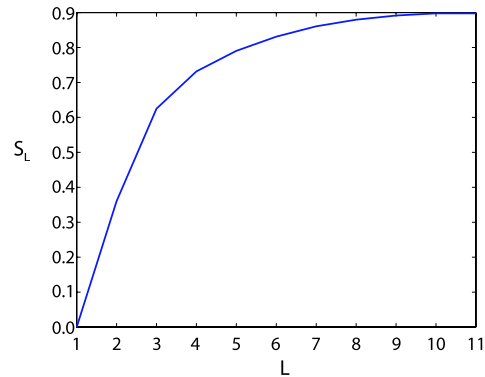}
\caption{\label{fig:XZZZ} Entropy of entanglement,
$S_L$, as a function of $L$ for a total of 19 spins. The plot shows
a logarithmic behavior that indicates criticality for
$\lambda_2=1$.}
\end{figure}
\end{center}

We can also determine numerically the central charge of the critical theory
and see if it corresponds to the same universality class. The central charge
can be obtained by studying the scaling behavior of the entropy of
entanglement \cite{Latorre}. The latter is given by the von Neumann entropy,
$S_L$, of the reduced density matrix, $\rho_L$. It can be calculated from
the density matrix of the original system, being in a pure state, where all
but $L$ contiguous spins are traced out. This indicates quantitatively the
degree of entanglement of the $L$ spins with the rest of the chain. Indeed,
we have
\begin{equation}
\begin{array}{rcl}
\rho_L & \equiv & \tr_{N-L} |\psi \rangle \langle\psi|, \\[\cdgap]
S_L & \equiv & \tr (\rho_L \log \rho_L),
\end{array}
\end{equation}
where $|\psi \rangle$ is the ground state of the system. For a critical
configuration and for large L we expect $\S_L \approx {c+\bar{c} \over 6}
\log L$, where $c$ is the central charge of the corresponding conformal
field theory and $\bar{c}$ is its complex conjugate. The central charge
uniquely corresponds to the critical exponents of the energy and of the
correlation length, $z$ and $\nu$, respectively. We know that for
non-critical chains $S_L$ should be saturated for large enough values of
$L$. This behavior is observed from the simulations when $\lambda_2\neq 1$.
On the other hand, when we are at $\lambda_2=B=1$ we obtain Figure
\ref{fig:XZZZ} that shows the expected logarithmic progression. By a
logarithmic fitting we can deduce that $c\approx 4/5$. Hence, our model is
in the same universality class as the three-state Potts model and it
corresponds to the critical exponents $z=1$ and $\nu=3/4$ in agreement with
the earlier findings.

\vspace{-1.4cm}
\section{The Full Hamiltonian}\label{sec:general}

We can now turn to the full Hamiltonian for arbitrary values of the coupling
parameters. For convenience we set $B = 1$ and vary only the interaction
couplings $\la_1$ and $\la_2$. Without loss of generality we can restrict on
the $\la_2>0$ half-plane as the case $\la_2<0$ is automatically obtained by
exchanging $\bm{\ua\leftarrow\!\!\!\!\rightarrow\da}$.

For large values of $\lambda_1$ or $\la_2$ it is possible to neglect
the transverse field $X = \sum_i \si_i^x$ and treat the system
classically. We find two asymptotes along which the ground state
should undergo a first order phase transition. By introducing $X$ as
a small perturbation one can predict how these curves behave when we
approach the origin $\lambda_1 =\lambda_2=0$. We find that along the
asymptote $\la_1/\la_2 = 3/2$ the classical ground state is highly
degenerate: it is in fact infinitely degenerate in the thermodynamic
limit. In the quantum regime, this gives rise to an intermediate
region between the two phases with spin-order 2 and 3, the size and
nature of which will be studied in this section.

\subsection{Perturbation Theory}

To understand the behavior of Hamiltonian (\ref{Ham}) we shall employ
perturbation theory. Surprisingly, the second order perturbation will give a
very good approximation to the numerical findings shown in Figure
\ref{fig:fidel}.

\subsubsection{Classical Regime} \label{sec:classical}

As a first step we take $\la_1,\la_2 \gg 1$, thus neglecting the $X$ term
the Hamiltonian reduces to
\begin{equation}
H \approx \lambda_1 \sum_i \si^z_{i}\si^z_{i+1} + \lambda_2 \sum_i
\si^z_{i}\si^z_{i+1}\si^z_{i+2} .
\end{equation}
This Hamiltonian has a classical behavior, so it can be solved exactly by
examining eigenstates of $\sigma_z^{\otimes N}$. There are three regions on
the $(\lambda_1,\lambda_2)$ plane of distinct ground state behavior.
Consider first $\lambda_1 < 0$, $\lambda_2
> 0$. The energy of the system is minimized uniquely by the state
\begin{equation}
\ket{A} = \ket{\downarrow\downarrow\downarrow\downarrow\downarrow
\downarrow \ldots \downarrow} .
\end{equation}
From the expectation values, $\langle A|  \sum_i \sigma^z_i
\sigma^z_{i+1} |A\rangle/N = 1$ and $\langle A| \sum_i \sigma^z_i
\sigma^z_{i+1} \sigma^z_{i+2} |A \rangle /N = -1$, one can calculate
the energy per site to be $E_A = \lambda_1 - \lambda_2$. Similarly
we take the cases of $\lambda_1 > 0$, $\lambda_2 \gg \lambda_1$ and
$\lambda_1>0$, $\lambda_2 \ll \lambda_1$, resulting in the following
ground states
\begin{equation}
|B\rangle  =  \ket{\uparrow\uparrow\downarrow\uparrow
\uparrow\downarrow \ldots \downarrow},
\end{equation}
with $ E_B =  -\1/3 \lambda_1 - \lambda_2$ and
\begin{equation}
|C\rangle = \ket{\ua\da\ua\da\ua\da \ldots \da},
\end{equation}
with $E_C = - \lambda_1$. Observe that these states are $3$ and $2$
periodic respectively giving rise to $Z_3$ and $Z_2$ symmetries. To
evaluate the position of the phase transitions we simply equate
energies in neighboring regions. In this way we obtain the
asymptotes $\lambda_1 = 0$ and ${\la_1} = \3/2\la_2$ that separate
the states of the system $\ket{A}$, $\ket{B}$ and $\ket{C}$ as seen
in Figure \ref{perturbation}.

Another classical region is obtained at $\la_1, \la_2  \ll 1$, where the
transverse field, $X$, dominates. In this region the ground state is given
by
\begin{equation}
\ket{D} = \ket{-}^{\otimes N},
\end{equation}
with $\ket{-} \equiv (\ket{\ua}-\ket{\da})/\sqrt{2}$ and an energy per site
of $E_D = -1$.

\subsubsection{Quantum Regime}

We now consider the transverse field as a perturbation to the theory, so we
rewrite our Hamiltonian as
\begin{equation}
H = \lambda_2 \sum_i(\lambda \sigma^x_i + \mu \sigma^z_i \sigma^z_{i+1} +
\sigma^z_i \sigma^z_{i+1}\sigma^z_{i+2}),
\end{equation}
where $\lambda \equiv {1 / \lambda_2}$ and $\mu\equiv {\la_1}/{\la_2}$.
Cyclic permutations of states $|B\rangle$ and $|C\rangle$ are also ground
states of the classical theory due to the translational invariance of the
Hamiltonian. Nevertheless, we need not employ degenerate perturbation theory
at this point since the permutations behave identically under the transverse
field, $X$. For this Hamiltonian the energies per site are
\begin{equation}
\tilde{E}_A = \lambda_1 - \lambda_2 - \frac{1}{\lambda_2} \frac{1}{6
- 4 \mu},
\end{equation}
\begin{equation}
\tilde{E}_B = - \left(\lambda_1 + \frac{\lambda_2}{3}\right) -
{\lambda_2} \left[\left(\frac{1}{9}\right) + \left(
\frac{1}{3}\right) \frac{1}{6 + 4\mu}\right].
\end{equation}
By inspection we see that the two energies are equal for $\mu=0$, which
coincides exactly with the classical solution. Hence, at $\O(\lambda^2)$,
the boundary between $\ket{A}$ and $\ket{B}$ remains unchanged, as seen in
Figure
\ref{perturbation}.

Let us turn to the perturbation treatment of the boundary between the
$\ket{D}$ and the $\ket{A}$ states. In fact, to the second order in the
couplings, $\la_1$ and $\la_2$, we find
\begin{equation}
\tilde{E}_D = -1-{\la_1^2 \over 4} - {\la_2^2 \over 6}.
\end{equation}
Equating $\tilde{E}_D = \tilde{E}_A$ we obtain the corresponding boundary to
be a correction of the classical one given by $\la_1 =\la_2-1$. The
computational solution of $\tilde{E}_D = \tilde{E}_A$ is given in Figure
\ref{perturbation}. It provides a line of second order phase transition
joining the Potts critical point and the Ising critical point that fits
firmly with the fidelity plot in Figure \ref{fig:fidel}.

\subsubsection{Competing Ground States}

We saw above that for $\mu = \frac{3}{2}$ the period 2 and period 3 phases
have the same ground state energy. Interestingly, on this asymptote the two
spin-orders can mix, giving rise in the thermodynamic limit to an infinitely
degenerate ground state. A perturbation theory around this area is
non-trivial as care has to be taken in the way the degeneracy is lifted.
Using a bosonization procedure similar to that presented in
\cite{Sachdev}, we will calculate the spectrum of the theory up to second order in the
transverse field. We will find a finite intermediate phase separating the
period 2 and period 3 gapped phases.

Let us present in detail the bosonization procedure necessary to develop the
perturbation theory. Consider two new types of bosons -- the $|3\rangle$,
created by the $t^\dagger$ operator, and $|2\rangle$, created by
$p^\dagger$. In the spin formalism, $|3\rangle$ corresponds to
$\ket{\ua\ua\da}$, while $|2\rangle$ is equivalent to $\ket{\ua\da}$. Thus,
\begin{equation}
\begin{array}{lcr}
|B\rangle = |33\ldots 3\rangle& \mbox{and} &|C\rangle = |22\ldots
2\rangle.
\end{array}
\end{equation}
All states that are composed of $2$s and $3$s are degenerate along the line
$\mu={3\over 2}$. Therefore, the region of the phase diagram that we want to
study is given by
\begin{equation}
0 < | \mu - 3/2 | \ll |\la_1|, |\la_2|. \label{eqn:regime}
\end{equation}
Since there are no first order contributions in the perturbation theory, the
region we want to study can be parameterized by
\begin{equation}
\mu = \3/2 + \sigma \lambda^2,
\end{equation}
for a dimensionless parameter $\sigma$.

The approach is to create a theory with $t$ and $p$ bosons that is
equivalent to the present one.  Let us first work with $t$ bosons. The
vacuum state has the form $|22 \ldots 2\rangle$, with no $3$s. To construct
the Hamiltonian we need the self energy of a $t$ boson, the interaction
energy and the hopping amplitude. For that, we calculate the energies per
site needed to add $22$, $33$ and $23$ bonds, given by $E_{22}$, $E_{33}$
and $E_{23}$, respectively. In particular,
\begin{equation}
|\underbrace{22\ldots 2}_{m-1}\,\rangle \rightarrow |\underbrace{22 \ldots
22}_{m}\,\rangle
\end{equation}
gives rise to the energy gap $E_{22}$, where $m$ is an integer. A similar
transformation gives $E_{33}$.  Furthermore,
\begin{equation}
|\underbrace{23 \ldots 23}_{2(m-1)}\,\rangle \rightarrow |\underbrace{23
\ldots 2323}_{(2m)}\,\rangle
\end{equation}
induces the energy gap $2 E_{23}$. The energy of each of these states can be
calculated perturbatively. Up to order $\O(\lambda^2)$, we obtain
\addtolength{\cdgap}{3mm}
\begin{equation}
\begin{array}{rcl}
E_{22} &=& -2\mu + \d\frac{\lambda^2}{2(1 - 2\mu)},\el E_{33}& =& -
(3 + \mu)  -  \d\frac{\lambda^2}{3}, \\[\cdgap]
E_{23}& =& -\d\3/2(1 + \mu)  - \d\frac{\lambda^2}{6}.
\end{array}
\end{equation}
Consider now the energy, $E_t$, of creating a single $t$ boson in a
background of $2$s, and the interaction energy, $E_{t, int}$, generated when
two $t$ bosons are brought to adjacent sites. The former can be evaluated by
considering the energy gap of the transformation
\begin{equation}
|\underbrace{22\ldots 222}_{5m}\,\rangle \rightarrow |\underbrace{23\ldots
23}_{4m}\,\rangle,
\end{equation}
which creates $2m$ bosons.  The states are chosen so that the lengths of the
underlying spin chains remain the same. For the interaction energy, the
relevant energy gap arises from
\begin{equation}
{|\underbrace{23\ldots 23}_{4m}\,\rangle} \rightarrow |\underbrace{2\ldots
2}_{2m}\underbrace{3\ldots 3}_{2m}\,\rangle.
\end{equation}
Hence these energies per boson are given by
\begin{equation}
\begin{array}{lcl}
E_t &=& 2 E_{23} - \d\frac{5}{2} E_{22}=
\lambda^2\left(\d\frac{7}{24} + 2 \sigma\right) ,\\[\cdgap]
E_{t, int} & =& 2 E_{23} - E_{22} - E_{33} = -
\d\frac{\lambda}{4}^2.
\end{array}
\end{equation}
Finally, we calculate the hopping amplitude, which is given by the
activation energy needed to perform the exchange
\[
\ket{23} \rightarrow \ket{32}.
\]
This has a value of $-{{\lambda^2} \over {6}}$. We are now in a position to
rewrite the Hamiltonian, which takes the following form
\begin{equation}
\begin{array}{rcl}
H & = & \d\frac{N}{2}\left(-2\mu + \d\frac{\lambda^2}{2(1 -
2\mu)}\right) \\[\cdgap]
&& - \lambda^2 \d\sum_l \left[\d\frac{1}{6} (t_{l+2}^\dagger t_l +
t_l^\dagger t_{l+2})\right.\\[\cdgap]
 & & - \, \left. t_l^\dagger t_l
\left(\d\frac{7}{24} + 2 \sigma\right) + \d\frac{1}{4}
t_{l+3}^\dagger t_l^\dagger t_{l+3} t_l\right] .
\end{array}
\end{equation}
The label $l$ is over the original indices, with a $t$ boson being centered
in the middle of the $3$ deformation, and the integer $N$ is the total
number of sites in the original spin formalism. We consider $\sigma
\gg 1$, which corresponds to the spin-order 2 phase. The ground state is the
vacuum state $\ket{22
\ldots 2}$, which is in agreement with original analysis. By employing
the Fourier transform of $t_l$ we obtain that the lowest excited energy
above the vacuum state is $\lambda^2\left(2 \sigma -
\frac{1}{24}\right)$. Thus at $\sigma = \frac{1}{48}$ a phase transition
occurs from period 2 to the intermediate region.

Following the same procedure we can find the boundary of the period 3 gapped
phase. Now the vacuum state is $|33 \ldots 3\rangle$ and the relevant
energies are with respect to the $p$ boson. We have
\begin{equation}
\begin{array}{lcl}
E_p & = & 2E_{23} - \d\frac{5}{3} E_{33} =  \lambda^2
\left(-\d\frac{4}{3} \sigma + \d\frac{2}{9}\right), \\
E_{p, int} & = & 2E_{23} - E_{22} - E_{33}= -\d\frac{\lambda}{4}^2,
\end{array}
\end{equation}
while the hopping amplitude is the same. In terms of these variables
the Hamiltonian takes the form
\begin{equation}
\begin{array}{rcl}
H & = & -\d\frac{N}{3}\left(3 + \mu + \d\frac{\lambda}{3}^2\right) \\[\cdgap]
&&- \lambda^2 \d\sum_l \left[\d\frac{1}{6}\left(p^\dagger_{l+3}p_l +
p_l^\dagger p_{l+3}\right)\right. \\[\cdgap]
& & \left.+\,p_l^\dagger p_l \left(\d\frac{4}{3} \sigma -
\d\frac{2}{9}\right) + \d{1 \over 4} \d p_{l+2}^\dagger p_l^\dagger
p_{l+2}p_l \right].
\end{array}
\end{equation}
An identical analysis is now necessary, by simply considering
$\sigma \ll -1$. The $p$ bosons cost large positive energy, making
the ground state the vacuum, as one would expect.  The first excited
state has an energy of $-\lambda^2\left(\frac{4}{3}\sigma +
\frac{7}{9}\right)$, so the boundary is at $\sigma = -\frac{7}{12}$.

\begin{figure}
\includegraphics[width=0.45\textwidth]{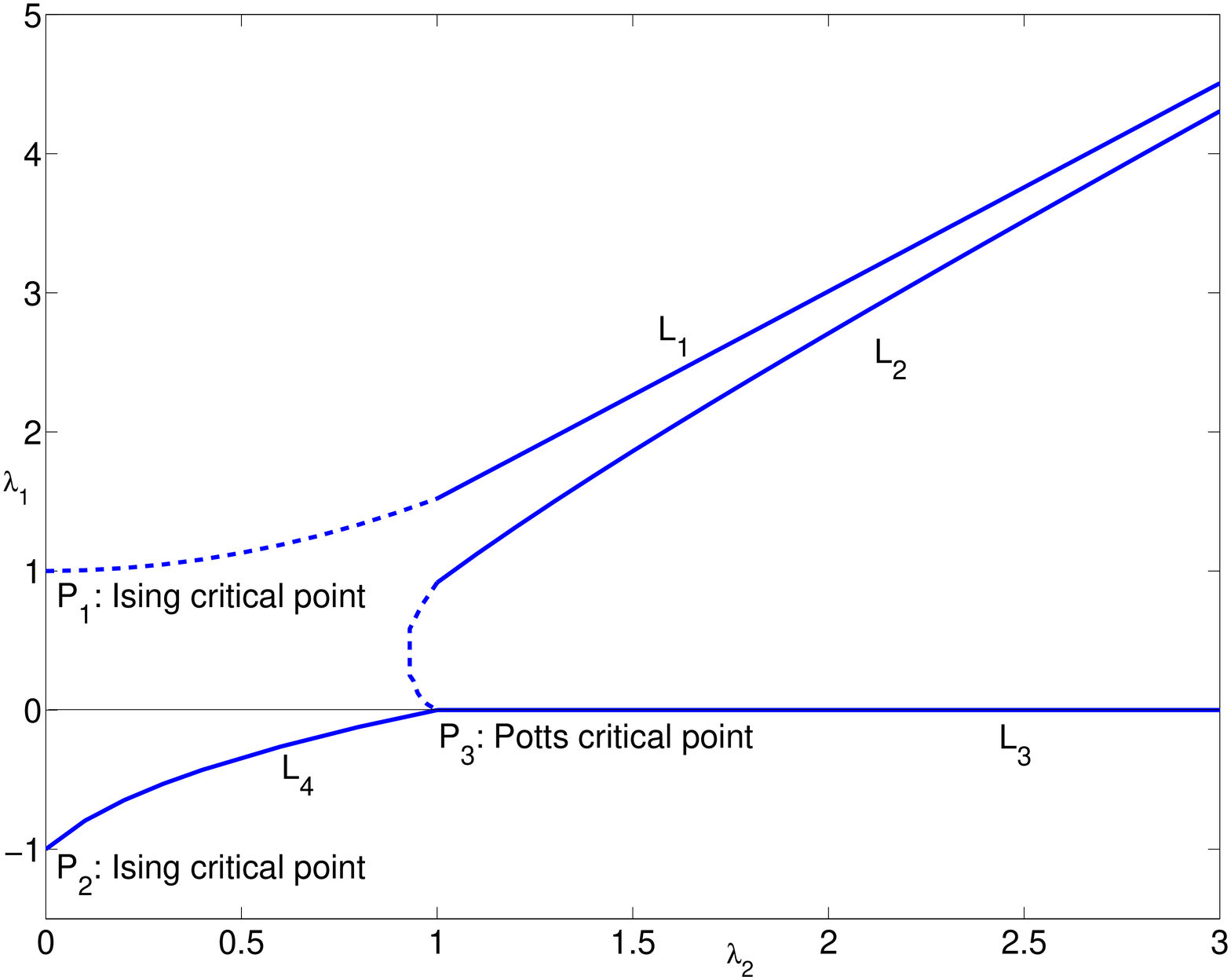}
\caption{\label{perturbation} An outline of the
perturbative results is
  given by the solid lines. Dotted lines are inserted by hand
  for completion. There
are three disjoint regions of different periodicity, separated by lines of
phase transitions, of both first ($L_3$) and second order ($L_1$,
  $L_2$ and $L_4$).}
\end{figure}

\vspace{1mm}We therefore have a prediction of the width of the
intermediate phase that is illustrated in Figure \ref{perturbation}.
This result is in perfect agreement with the fidelity plot in
Figure \ref{fig:fidel}.

\subsection{Criticality Behavior}

At this point we can discuss the criticality behavior of our system.
Apart from the Ising critical points, P$_1$ and P$_2$, and the
three-state Potts model critical point, P$_3$, we can identify the
critical behavior of the lines, L$_1$, L$_2$, L$_3$ and L$_4$, seen
in Figure \ref{perturbation}, where our results of the perturbation
treatment are summarized.

Specifically, we obtain first order critical behavior for $\la_1 = 0$ and
$\la_2>1$ and second order behaviors for the rest of the curves. The
perturbation theory shows that the line, L$_3$, acquires no second order
contribution indicating its first order nature. This is also supported by a
numerical study of the energy gap for a chain of 18 spins.  Indeed, we find
that its minima lie along lines similar to the ones in Figure
\ref{perturbation}. Furthermore, there is an actual energy crossing along
the first order line while all the other critical points remain gapped.

Studying the critical exponents along the line L$_4$ we find that in moving
from P$_3$ to P$_2$ along the line of criticality, the exponent $z$
decreases to zero and becomes discontinuous at the point P$_2$. Moving in
the other direction along L$_2$ yields ever increasing values of $z$. Hence,
in view of Eqn. (\ref{eqn:Delta}), we confirm that as the transverse field
becomes insignificant, the phase transitions return to first order behavior
with an actual energy level crossing. Along the upper asymptote, L$_1$, we
observe similar behavior. The critical exponent is large in configurations
far away from the origin and decreases continuously down to $z=1/2$ at the
Ising critical point P$_1$.

These results present a rich variety of phase transition behavior. From
Figures \ref{fig:fidel} and \ref{perturbation} we see that the second order
perturbation theory gives a good picture of the ground state of the full
Hamiltonian in most of the parametric region. The only part of the
$(\la_1,\la_2)$ plane that is not yet understood is the strip bounded by
the lines, $L_1$ and $L_2$. Probing the physics of this region is the
subject of the next subsection.

\subsection{Incommensurate Phase}

It becomes apparent from the previous that perturbation theory cannot give
reliable information about the physics between $L_1$ and $L_2$. To study
that region we resort to the Landau-Ginzburg approximation \cite{QPT}. As we
shall see, the resulting theory corresponds to the chiral clock model
\cite{Ostlund}, which predicts non-zero chirality for our model. This also
indicates the presence of a gapless incommensurate phase between $L_1$ and
$L_2$. For a finite chain we find strong evidence for the presence of this
phase using exact numerical diagonalization.

To gain a further insight into the nature of the ground states of our model
we introduce the order parameter
\cite{Sachdev}
\begin{equation}
\Psi_p = \sum_j e^{2\pi i j / p} \ket{\da}_j\bra{\da},
\end{equation}
where the operator $\ket{\da}_j\bra{\da}$ is the projector onto the down
state of the $j^\text{th}$ site. The periodicity of a state is revealed by
the expectation value of $\Psi_p$: it is maximal for $p$-periodic states,
and minimal (or zero) otherwise. In particular, the wave order parameter,
$\Psi_3$, is a complex operator that detects the period 3 states. We wish to
use $\Psi_3$ as an order parameter to probe the behavior of our model
between the boundaries $L_1$ and $L_2$. With this in mind we construct a
continuum quantum field theory with an action that has the same symmetries
as $\lan
\Psi_3 \ran$ does in the $Z_3$ region.

It is easy to verify that $\lan\Psi_3 \ran$ is invariant under translations
$\Psi_3
\rightarrow e^{2\pi i l/3}\Psi_3$, for integer $l$. A more intriguing
symmetry is given when simultaneous spatial reflections and complex
conjugation are performed, $x \rightarrow -x, \Psi_3
\rightarrow \Psi_3^*$ \cite{Huse}. Finally, $\lan \Psi_3 \ran$ is
invariant under time reversal, $\tau \rightarrow - \tau$. According to these
symmetries, we can write down the corresponding effective field theory,
given by,
\begin{equation}
\begin{array}{rcl}
\S_3 &=& \d\int dx d\tau \left[i\alpha \Psi_3^*
\p_x \Psi_3 + \text{c.c.} + \left|\p_\tau\Psi_3\right|^2\right. \\
& + &v^2\left|\p_x\Psi_3\right|^2+\left. r\left|\Psi_3\right|^2 +
v\Psi_3^3 + \text{c.c.} + \ldots\right].
\end{array}
\end{equation}
The parameters $\al, v$ and $r$ are non-universal functions of the
original $\la_1$ and $\la_2$.  It has been shown \cite{Baxter} that
this model has a critical point at $\al = r= 0$ that belongs in the
same universality class as the point P$_3$. We may therefore
identify them.

The term $\al\Psi_3^*\p_x\Psi_3$ breaks chiral symmetry when
$\Psi_3$ is complex. The resulting theory has been studied by
Ostlund \cite{Ostlund} and it has been shown that it belongs in the
same universality class as the chiral clock model. Hence, it
suggests the presence of a chiral phase for non-zero values of
$\al$. Furthermore this model predicts that this phase is gapless
and incommensurate \cite{Huse,Cardy}. By analogy we conjecture that
our model has such a phase away from $P_3$ and we shall investigate
it numerically.

\begin{figure}[t]
\includegraphics[width=0.45\textwidth]{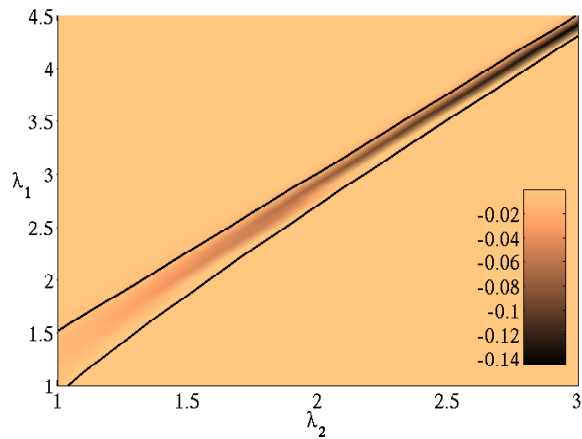}
\caption{\label{fig:excite} Surface plot of $\langle
\Psi_{18/7} \rangle$. There is a narrow band of excitation within
the intermediate region predicted by perturbation theory, giving
strong indication for the presence of gapless incommensurate phase.}
\end{figure}

\begin{figure}[t]
\includegraphics[width=0.45\textwidth]{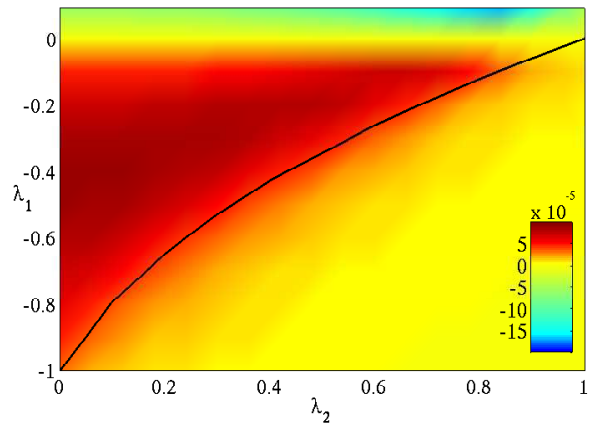}
\caption{\label{fig:diff} This plot shows $\langle
\Psi_{18/5} - \Psi_{18/7}\rangle$ over $\lambda_1<0, \lambda_2 < 1$.
There is a clear region, bounded by the predictions of perturbation
theory, where $|\psi\rangle$ responds best to excitations of
spin-order $p=18/5>3$, indicating, in tandem with Figure
\ref{fig:excite}, a theory with $p<3$ above $P_3$ and $p>3$ below.}
\end{figure}

The gapless and incommensurate characteristics of this phase
indicate that $\lan \Psi_p \ran$ can build a non-zero value by
applying the appropriate perturbation. Hence, we add to the
Hamiltonian a small periodic potential of the form
\begin{equation}
V = 10^{-4} \sum_j e^{2\pi i j / p} \ket{\da}_j \bra{\da}
\end{equation}
and we plot $\lan \Psi_p \ran$ for several $p$ as a function of $\la_1$ and
$\la_2$. Indeed, for $p<3$, we find that $\lan \Psi_p \ran \neq 0$ between
$L_1$ and $L_2$ and deduce that $\al \neq 0$ there (see Figure
\ref{fig:excite}).  A naive analysis of our field theory in which we treat
$\Psi_p$ as a fluctuation of the uniform solution indicates that $\al$ may
be proportional to $2\pi/p - 2\pi/3$. This suggests a region of $\alpha\neq
0$, with opposite sign corresponding to $p>3$, in which a second chiral
phase exists on the other side of $P_3$. We discover numerical evidence for
this in Figure
\ref{fig:diff} when $\la_1 < 0$ and $\la_2 < 1$, but its effect is
of significantly lower order than in Figure \ref{fig:excite}.


\begin{figure}[h]
\includegraphics[width=0.34\textwidth]{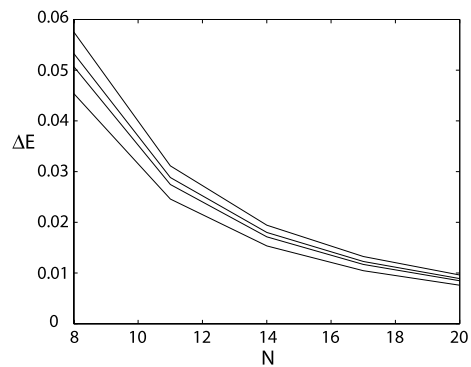}
\caption{\label{fig:degen} The energy gap, $\Delta E$, per site
between the ground state and first excited state tends exponentially to zero
as the total length of the chain increases. In the thermodynamical limit we
expect to have zero energy gap. The lines correspond to different points
along the $\la_1 /\la_2=3/2$ asymptote, ordered from top to bottom the
further away the points are from the origin.}
\end{figure}

Along with the gapless incommensurate properties of this phase we
would like to investigate its chiral nature. Note that, on the one
hand, the chirality operator, given by $\chi_i \equiv
\vec{\sigma}_i\cdot \vec{\sigma}_{i+1} \times \vec{\sigma}_{i+2}$,
is an imaginary hermitian operator. On the other hand the Hamiltonian
(\ref{Ham}) is real. Thus, if the Hamiltonian possesses a ground state with
non-zero expectation value $\langle\chi\rangle$, that ground state should be
multidegenerate. If this is the case in the thermodynamic limit we would
like to see if it is also approximated in the finite case. By numerical
diagonalization of the finite-size Hamiltonian, we can obtain the lowest
energy gap, $\Delta E$, for values of the couplings $\la_1$ and $\la_2$
between the $L_1$ and $L_2$ lines. In Figure
\ref{fig:degen} we can clearly observe an exponential damping of $\Delta E$
as a function of the length of the spin chain. This allows us to assume that
in the thermodynamic limit the spectrum will eventually become degenerate.

\section{Conclusion}\label{Conclusion}

The present spin model approximates the behavior of the Mott insulator of
two species of atoms when a triangular ladder geometry is realized. The spin
interpretation, which makes the study of the model simpler and more
intuitive, is valid when there is only one atom per lattice site, i.e. at
the regime of strong collisional couplings. With this condition satisfied
the criticality behavior of the spin model straightforwardly represents the
behavior of the Mott insulator. We can argue that non-zero chirality in the
spin model corresponds to a ground state with persistent currents
\cite{Wang}. This suggests a counterflow of the two different atomic species
that could be measured by using atomic spatial correlations
\cite{Grondal}.

To summarize, in this article, we have generalized the one-dimensional Ising
model, $ZZ$, to include an additional triple $\sigma^z$ interaction, $ZZZ$,
in the presence of a transverse magnetic field. A rich criticality behavior
has been revealed when the coupling of $ZZ$ and $ZZZ$ are varied. The phases
of the ground state are identified according to their periodic structure. In
particular, a complex order parameter related to the states with period
three has revealed a chiral, gapped, incommensurate phase. In
\cite{Sachdev} a thorough study of this phase in a one-dimensional
hard boson model exhibiting similar behavior was performed. An in
depth study of the incommensurate phase will be the subject of a
future publication.

\acknowledgements

This work was supported by the Royal Society.
\bibliographystyle{asprev}

\begin{thebibliography}{1}


\bibitem{Jaks98}
D. Jaksch, C. Bruder, J.I. Cirac, C.W. Gardiner, and P. Zoller,
\href{http://prola.aps.org/abstract/PRL/v81/i15/p3108_1}{Phys. Rev.
Lett. {\bf 81} 3108} (1998).

\bibitem{Raithel} A. Kastberg, W. D. Phillips, S. L. Rolston,
  R. J. C. Spreeuw, and P. S. Jessen, \href{http://prola.aps.org/abstract/PRL/v74/i9/p1542_1}{Phys. Rev. Lett. {\bf 74},
  1542}
  (1995); G. Raithel, W. D. Phillips, and S. L. Rolston,
  \href{http://prola.aps.org/abstract/PRL/v81/i17/p3615_1}{Phys. Rev. Lett. {\bf 81}, 3615} (1998).

\bibitem{Mandel1}
M. Greiner, O. Mandel, T. Esslinger, T. W. H\"{a}nch, and I. Bloch,
\href{http://www.nature.com/cgi-taf/DynaPage.taf?file=/nature/journal/v415/n6867/abs/415039a_fs.html}
{Nature {\bf 415}, 39} (2002); M. Greiner, O. Mandel, T. W. Heanch,
and I. Bloch,
\href{http://www.nature.com/cgi-taf/DynaPage.taf?file=/nature/journal/v419/n6902/abs/nature00968_fs.html}{Nature
{\bf 419}, 51} (2002).

\bibitem{Mandel3}
O. Mandel, M. Greiner, A. Widera, T. Rom, T. W. Heanch, and I.
Bloch,
\href{http://www.nature.com/cgi-taf/DynaPage.taf?file=/nature/journal/v425/n6961/abs/nature02008_fs.html}{Nature
{\bf 425}, 937} (2003).

\bibitem{Deutsch}
I. H. Deutsch, G. K. Brennen and P. S. Jessen, Forsch. der Phys.
(2000),
\href{http://arxiv.org/abs/quant-ph/0003022}{quant-ph/0003022}.

\bibitem{Jaksch}
D. Jaksch, to appear in Contemporary Physics,
\href{http://arxiv.org/abs/quant-ph/0407048}{quant-ph/0407048}.

\bibitem{Ripoll}
J. J. Garcia-Ripoll and I. J. Cirac,
\href{http://www.journals.royalsoc.ac.uk/link.asp?id=2pr03a1c5e0f383n}{Phil.
Trans. R. Soc. Lond. A 361, 1537-1548} (2003).

\bibitem{Kuklov}
A. Kuklov, N. Prokof'ev, and B. Svistunov Phys. Rev. Lett. {\bf 92}, 050402
(2004).

\bibitem{Belen} B. Paredes, A. Widera, V. Murg, O. Mandel, S. F\"olling,
I. Cirac, G. V. Shlyapnikov, T. W. H\"{a}nsch, and I. Bloch,
\href{http://www.nature.com/cgi-taf/DynaPage.taf?file=/nature/journal/v429/n6989/full/nature02530_fs.html}{Nature
{\bf 429}, 277} (2004).

\bibitem{Kukl}
A.B. Kuklov, and B.V. Svistunov,
\href{http://link.aps.org/abstract/PRL/v90/e100401}{Phys. Rev. Lett.
{\bf 90}, 100401} (2003).

\bibitem{Jask03}
D. Jaksch, and P. Zoller,
\href{http://www.iop.org/EJ/abstract/1367-2630/5/1/356}{New Journal
Phys. {\bf 5}, 56.1} (2003).

\bibitem{Duan}
L.M. Duan, E. Demler, and M. D. Lukin,
\href{http://link.aps.org/abstract/PRL/v91/e090402}{Phys. Rev. Lett.
{\bf 91}, 090402} (2003).

\bibitem{Pachos}
J. K. Pachos, and E. Rico,
\href{http://link.aps.org/abstract/PRA/v70/e053620}{Phys. Rev. A 70,
053620} (2004).

\bibitem{Sachdev}
P. Fendley, K. Sengupta, and S. Sachdev,
\href{http://link.aps.org/abstract/PRB/v69/e075106}{Phys. Rev. B.
{\bf69}, 75106} (2004).

\bibitem{Nielsen}
C. M. Dawson, and M. A. Nielsen, Phys. Rev. A {\bf 69}, 052316 (2004).

\bibitem{Wu}
F. Y. Wu, Rev. Mod. Phys. {\bf 54}, 235 (1982).

\bibitem{Pachos1}
A. Kay, D. K. K. Lee, J. K. Pachos, M. B. Plenio, M. E. Reuter and E. Rico,
Optics and Spectroscopy {\bf 99}, 355 (2005),
\href{http://arxiv.org/quant-ph/0407121}{quant-ph/0407121}.

\bibitem{Alcaraz}
For a study of the corresponding classical model see: F. C. Alcaraz
and J. C. Xavier, J. Phys. A: Math. Gen. {\bf 32} 2041 (1999);
D. W. Wood and H. P. Griffiths, J. Phys. C: Solid State Phys. {\bf 5}
L253 (1972).

\bibitem{Penson}
K. A. Penson, R. Jullien, and P. Pfeuty, Phys. Rev. B {\bf 26}, 6334
(1982); K. A. Penson, J. M. Debierre, and L. Turban, Phys. Rev. B {\bf
  37}, 7884 (1988).

\bibitem{Igloi}
F. Igloi, J. Phys. A {\bf 20}, 5319 (1987); Phys. Rev. B {\bf 40},
2362 (1989); J. C. Angles d'Auriac, and F. Ingloi, Phys. Rev. E {\bf
  58}, 241 (1998).

\bibitem{Turban}
Lo\"{i}c Turban,
\href{http://www.iop.org/EJ/S/0/21257/Jvt0nUNf.i2w20Vnqg7dXQ/article/0022-3719/15/4/006/jcv15i4pL65.pdf}{J.
Phys. C {\bf 15} L65-L68} (1982); K. A. Penson, R. Jullien, and P.
Pfeuty, Phys. Rev. B {\bf 26}, 6334 (1982).

\bibitem{Goldenfield}
N. Goldenfield, {\it Lectures on phase transitions and the
renormilisation group}, Addison-Wesley (1992).

\bibitem{Latorre}
J. I. Latorre, E. Rico, and G. Vidal, QIC {\bf 4}, 48 (2004).

\bibitem{QPT} S. Sachdev, {\it Quantum Phase Transitions}, Cambridge
University Press (2001).

\bibitem{Ostlund} S. Ostlund, Phys. Rev. B {\bf 24}, 398 (1981).

\bibitem{Huse} D. A. Huse and M. E. Fisher, Phys. Rev. Lett. {\bf
49}, 793 (1982).

\bibitem{Baxter}
R. J. Baxter, J. Phys. A {\bf 13}, L61 (1980); J. Stat. Phys. {\bf 26},427
(1981).

\bibitem{Cardy}
J. L. Cardy, Nucl. Phys. B {\bf 389}, 577 (1993).

\bibitem{Wang}
W. Zhuo, X. Wang, and Y. Wang,
\href{http://arxiv.org/pdf/cond-mat/0501693}{cond-mat/0501693}.

\bibitem{Grondal}
J. P. Grondalski, P. M. Alsing, and I. H. Deutsch, {Opt. Exp. {\bf 5}, 249},
(1999).

\end{thebibliography}

\end{document}